\magnification \magstep1
\raggedbottom
\openup 2\jot 
\voffset6truemm
\centerline {\bf CASIMIR ENERGY IN NON-COVARIANT GAUGES}
\vskip 1cm
\noindent
Giampiero Esposito$^{1,2}$, Alexander Yu. Kamenshchik$^{3,4}$
and K. Kirsten$^{5}$
\vskip 0.3cm
\noindent
${ }^{1}$Istituto Nazionale di Fisica Nucleare, Sezione di Napoli,
Complesso Universitario di Monte S. Angelo, Via Cintia, Edificio N',
80126 Napoli, Italy
\vskip 0.3cm
\noindent
${ }^{2}$Universit\`a degli Studi di Napoli Federico II,
Dipartimento di Scienze Fisiche,
Complesso Universitario di Monte S. Angelo, Via Cintia, Edificio N',
80126 Napoli, Italy
\vskip 0.3cm
\noindent
${ }^{3}$L.D. Landau Institute for Theoretical Physics, Russian
Academy of Sciences, Kosygina Str. 2, Moscow 117334, Russia
\vskip 0.3cm
\noindent
${ }^{4}$Landau Network-Centro Volta, Villa Olmo, Via Cantoni 1,
22100 Como, Italy
\vskip 0.3cm
\noindent
${ }^{5}$The University of Manchester, Department of Physics and
Astronomy, Theory Group, Schuster Laboratory, Oxford Road,
Manchester M13 9PL, England
\vskip 1cm
\noindent
{\bf Abstract}.
The zero-point energy of a conducting spherical shell is
studied by imposing the axial gauge via path-integral methods,
with boundary conditions on the electromagnetic potential
and ghost fields. The coupled modes are then found to be the
temporal and longitudinal modes for the Maxwell field. The
resulting system can be decoupled by studying a fourth-order
differential equation with boundary conditions on longitudinal
modes and their second derivatives. Complete agreement is found
with a previous path-integral analysis in the Lorenz gauge,
and with Boyer's value. This investigation leads to a better
understanding of how gauge independence is achieved in quantum
field theory on backgrounds with boundary.
\vskip 100cm
The quantization programme of fundamental interactions in 
non-covariant gauges has been always of great importance, as is
well known from the work in Refs. [1,2]. On the other hand, recent
progress in the measurement of Casimir energies [3--6] motivates new
theoretical efforts, including the comparison of covariant and
non-covariant gauges for the evaluation of Casimir energies.

We here rely entirely on the path-integral analysis of Ref. [7],
and avoid review for length reasons. Recall therefore that, in the 
case of a spherical shell of radius $R$, the axial gauge forces 
normal modes of the potential to vanish everywhere, jointly with
ghost modes. One is then left with transverse modes 
$T_{l}(r)=\sqrt{\pi /2}\sqrt{r}I_{l+1/2}(Mr)$ which obey homogeneous
Dirichlet conditions at $r=R$, while temporal and longitudinal
modes (denoted by $a_{l}$ and $c_{l}$, respectively), obey the
coupled system [7]
$$
\left[{d^{2}\over dr^{2}}+{2\over r}{d\over dr}-{l(l+1)\over r^{2}}
\right]a_{l}-{Ml(l+1)\over r^{2}}c_{l}=0,
\eqno (1)
$$
$$
\left[{d^{2}\over dr^{2}}-M^{2}\right]c_{l}-Ma_{l}=0.
\eqno (2)
$$
First, we now point out that this system can be decoupled if $a_{l}$
is expressed from Eq. (2) and the result is inserted into Eq. (1). 
On setting $y \equiv Mr$, this leads to the fourth-order equation
$$
\left[{d^{4}\over dy^{4}}+{2\over y}{d^{3}\over dy^{3}}
-\left(1+{l(l+1)\over y^{2}}\right){d^{2}\over dy^{2}}
-{2\over y}{d\over dy}\right]c_{l}(y)=0.
\eqno (3)
$$
Moreover, we know from Ref. [7] that, at $r=R$, both $a_{l}$ and 
$c_{l}$ should vanish. By virtue of Eq. (2) this implies that,
at $r=R$, both $c_{l}$ and its second derivative should vanish. 
The origin $r=0$ is a regular singular point of our boundary-value
problem, and hence we require that both $a_{l}$ and $c_{l}$ should
vanish therein. By virtue of Eq. (2), this implies that also the
second derivative of $c_{l}$ with respect to $y$ vanishes at $y=0$.
Now we set $F_{l}(y) \equiv c_{l}'(y)$, and cast Eq. (3) in the form
$$
{1\over y^{2}}{d\over dy}\left[y^{2}\left({d^{2}\over dy^{2}}
-\left(1+{l(l+1)\over y^{2}}\right)\right)F_{l}(y)\right]=0.
\eqno (4)
$$
On denoting by $C$ a constant, the desired $F_{l}(y)$ solves therefore
the inhomogeneous equation
$$
\left[{d^{2}\over dy^{2}}-\left(1+{l(l+1)\over y^{2}}\right)
\right]F_{l}(y)={C\over y^{2}},
\eqno (5)
$$
whose regular solution reads ($d$ being a constant)
$$
F_{l}(y)=CF_{sp}(y)+d \sqrt{\pi /2}\sqrt{y}I_{l+1/2}(y),
\eqno (6)
$$
where $F_{sp}$ is chosen to be a particular 
solution of Eq. (5) with $C=1$ whose first
derivative vanishes at $0$ and at $b \equiv MR$. The vanishing of $c_{l}$
at the boundary fixes the relation between $C$ and $d$, i.e.
$$
C=-d{\int_{0}^{b}dz \; \sqrt{\pi /2}\sqrt{z}I_{l+1/2}(z)
\over \int_{0}^{b} dz \; F_{sp}(z)}.
\eqno (7)
$$
Eventually, the vanishing of $c_{l}''$ at the boundary leads to
$$
I_{l+1/2}'+{1\over 2b}I_{l+1/2}(b)=0,
\eqno (8)
$$
because, by construction, $CF_{sp}$ is a particular solution of Eq. (5)
whose first derivative vanishes at the boundary. Equation (8), jointly
with the vanishing at the boundary of transverse modes, yields the same
set of eigenvalue conditions, with the same degeneracies, found for the
interior problem in the Lorenz gauge in Ref. [7]. Thus, complete agreement
[8] with Boyer's value [9] for the Casimir energy is recovered.

Although non-covariant gauges break relativistic covariance, they make it
possible to decouple Faddeev--Popov ghosts from the gauge field. Thus,
ghost diagrams do not contribute to cross-sections and need not be evaluated,
and this property has been regarded as the main advantage of non-covariant
gauges [2]. In the case of Casimir energies, the ghost field is forced to
vanish everywhere by virtue of the boundary conditions appropriate for the
axial gauge [7], and the analysis of temporal and longitudinal modes,
although rather involved, has been proved to lead to the same Casimir
energy [9] for the interior problem in the Lorenz gauge [7]. The particular
solution of the inhomogeneous equation (5) plays a non-trivial role in
ensuring that, despite some technical points, the resulting Casimir energy
is the same as in the Lorenz gauge, hence proving explicitly the equivalence
of a covariant and a non-covariant gauge for a conducting spherical shell.
A better theoretical understanding of gauge independence in quantum field
theory has been therefore gained, after the encouraging experimental progress
of recent years in the measurement of Casimir forces [3--6].
\vskip 0.3cm
\noindent
{\bf Acknowledgments}. Klaus Kirsten thanks Stuart Dowker for interesting
discussions. The work of Klaus Kirsten has been supported by EPSRC, 
grant no GR/M08714.
\vskip 0.3cm
\leftline {\bf References}
\vskip 0.3cm
\noindent
\item {[1]}
G. Leibbrandt, {\it Rev. Mod. Phys.} {\bf 59}, 1067 (1987).
\item {[2]}
G. Leibbrandt, {\it Noncovariant Gauges} (World Scientific,
Singapore, 1994).
\item {[3]}
S. K. Lamoreaux, {\it Phys. Rev. Lett.} {\bf 78}, 5 (1997).
\item {[4]}
U. Mohideen and A. Roy, {\it Phys. Rev. Lett.} {\bf 81},
4549 (1998).
\item {[5]}
A. Roy and U. Mohideen, {\it Phys. Rev. Lett.} {\bf 82},
4380 (1999).
\item {[6]}
A. Roy, C. Y. Liu and U. Mohideen, {\it Phys. Rev.} 
{\bf D 60}, 111101 (1999).
\item {[7]}
G. Esposito, A. Yu. Kamenshchik and K. Kirsten, 
{\it Int. J. Mod. Phys.} {\bf A 14}, 281 (1999).
\item {[8]}
G. Esposito, A. Yu. Kamenshchik and K. Kirsten, ``Casimir Energy
in the Axial Gauge'' (DSF preprint 2000/21, HEP-TH 0006220).
\item {[9]}
T. H. Boyer, {\it Phys. Rev.} {\bf 174}, 1764 (1968).

\bye